\documentclass[twocolumn,aps]{revtex4}
\usepackage{graphicx}
\usepackage{amsbsy}

\begin{document}

\title{Mixed partial-wave scattering with spin-orbit coupling\\
 and validity of pseudo-potentials}
\author{Xiaoling Cui}
\affiliation{Institute for Advanced Study, Tsinghua University, Beijing, 100084 \\
Department of Physics, Ohio State University, Columbus, OH 43210 }
\date{{\small \today}}
\begin{abstract}
We present exact solutions of two-body problem for spin-$1/2$
fermions with isotropic spin-orbit(SO) coupling and interacting with
an arbitrary short-range potential. We find that in each
partial-wave scattering channel, the parametrization of two-body
wavefunction at short inter-particle distance depends on the
scattering amplitudes of all channels. This reveals the mixed
partial-wave scattering induced by SO couplings. By comparing with
results from a square-well potential, we investigate the validity of
original pseudo-potential models in the presence of SO coupling. We
find the s-wave pseudo-potential provides a good approximation for
low-energy solutions near s-wave resonances, given the length scale
of SO coupling much longer than the potential range. However, near
p-wave resonance the p-wave pseudo-potential gives low-energy
solutions that are qualitatively different from exact ones, based on
which we conclude that the p-wave model can not be applied to the
fermion system if the SO coupling strength is larger or comparable
to the Fermi momentum.
\end{abstract}

\maketitle

\section{Introduction}

Two-body problem takes the most fundamental place in the process of
exploring and understanding many-body properties. In particular,
two-body solutions determine the essential interaction parameter in
the microscopic many-body Hamiltonian. In the field of dilute
ultracold atoms, the two-body interaction is generally formulated by
the zero-range pseudo-potential, provided that it produces the same
asymptotic two-body wavefunction at length scale much shorter than
the mean inter-particle distance but longer than the range of
realistic potential. The generalized pseudo-potentials for all
partial-waves were first derived by Huang and Yang\cite{LHY}, and
then improved later by Stock {\it et al}\cite{Stock05}. So far the
most popular pseudo-potential is in s-wave channel described by a
single s-wave scattering length, which can be improved by including
the energy-dependence in a self-consistent way\cite{Stock05,
E-dependence}. Another popular one is the p-wave pseudo-potential,
described by the p-wave scattering volume which generally has strong
energy-dependence\cite{Gurarie, Yip}.

In view of the great success when applying pseudo-potential models
to the homogenous or trapped atomic gases, it is generally believed
that this model will equally apply to other configurations, such as
in the presence of spin-orbit(SO) coupling. Recently, by
sophisticated manipulations of laser field and magnetic field, the
NIST group has successfully realized an optically synthesized
magnetic field for ultracold neutral atoms\cite{expe_soc}. As a
result, an effective SO coupling is generated in the system along
one direction. Subsequently there are several theoretical proposals
to realize the symmetric Rashba SO coupling\cite{Rashba}, and it is
conceivable that an arbitrary type SO coupling could be achieved in
future experiments. As usual, all existing theoretical studies about
the SO coupled system are carried out in the framework of s-wave
pseudo-potential, i.e., using the s-wave scattering length as that
without SO coupling(see recent review \cite{HZ_review}). Based on
this model, the most remarkable effect of symmetric SO coupling is
to support a two-body bound state with an arbitrarily weak
interaction, due to the modified low-energy density
of state\cite{2-body}. 

Although pseudo-potentials have been justified under confinement
potentials\cite{Stock05, E-dependence, Yip}, it is not obvious that
it is still robust under the single-particle potential as special as
SO couplings. In the two-body scattering process, trapping
potentials and SO couplings have the same effect in mixing different
partial-waves, either due to the trap anisotropy\cite{mix_trap,
Nishida}, or due to the intermediate coupling with spin sector.
However, unlike the trapping potentials, which generally contribute
a trivial constant potential as two particles get close, the SO
coupling intrinsically affects the kinetic term and thus still mix
all partial-waves for the two-body wavefunctions at short
inter-particle distance. This non-trivial effect is expected to have
important influence on the validity of original pseudo-potentials in
the presence of SO coupling. For instance, an obvious deficiency of
original s-wave pseudo-potential is that this model can predict
arbitrarily deep bound state with the binding energy scaled in terms
of the SO coupling strength\cite{2-body}; however, under a
square-well (attractive) interaction potential the true binding
energy must be lower-bounded by the potential depth. Moreover, this
discrepancy can not be amended by taking into account the
energy-dependence of s-wave scattering length, as we shall show
later in Section IV.

In this paper, we make efforts to exactly solve the two-body problem
with SO coupling for a general short-range interaction potential,
without resorting to pseudo-potential models. For simplicity but
without the loss of essence, we have chosen the isotropic SO
coupling and studied in the subspace where only s-wave and p-wave
scatterings are relevant. 
We show that the short-range parametrization of the wavefunction in
each partial-wave channel will additionally rely on the scattering
amplitude of another partial-wave channel, which reflects the mixed
scattering between different orbital channels induced by SO
coupling. The exact form of wavefunction obtained above allows us to
solve the two-body problem under a square-well interacting
potential. By comparing with results from s-wave or p-wave
pseudo-potentials, we address the validity of the latter in the
presence of isotropic SO coupling. We find the s-wave
pseudo-potential provides a good approximation for the low-energy
scattering state and bound state solutions near s-wave resonance,
with the correction depending on the strength of SO coupling, the
finite range of the potential and contributions from p-wave channel.
However, near p-wave resonance, using the p-wave pseudo-potential
alone will lead to results that are qualitatively different from
exact solutions from the square-well potential. We conclude that the
p-wave pseudo-potential can not be applied to fermion system if the
SO coupling strength is larger or comparable to the Fermi momentum.
We shall address the underlying
reasons for these results. 

The rest of the paper is organized as follows. In section II, we
present the exact solution for two spin-$1/2$ fermions under a
general short-range interaction potential and with isotropic SO
coupling. In section III we reduce the exact solutions to the
framework of original s-wave and p-wave pseudo-potentials. In
section IV we present the numerical results for two-body problem
under the square-well potential, from which we address the validity
of s-wave and p-wave pseudo-potential models. We summarize the paper
in the last section.

\section{Two-body problem with isotropic spin-orbit coupling}

In this section we shall solve the two-body scattering problem for a
special case of isotropic SO coupling. Assuming a general form of
short-range interaction potential (See Eq.\ref{U_P}), we obtain the
wavefunction of scattering state(Eq.\ref{psi_01_super2}) and analyze
its long-range and short-range asymptotic behaviors. Particularly we
show that its short-range behavior is parameterized by scattering
amplitudes in all relevant partial-wave channels
(Eqs.\ref{short_psi0_f},\ref{short_psi1_f}). Finally we present the
bound state solutions which can be deduced from scattering state
solutions via Eq.\ref{bs}.

\begin{figure}[hbtp]
\includegraphics[height=3.8cm,width=8.3cm]{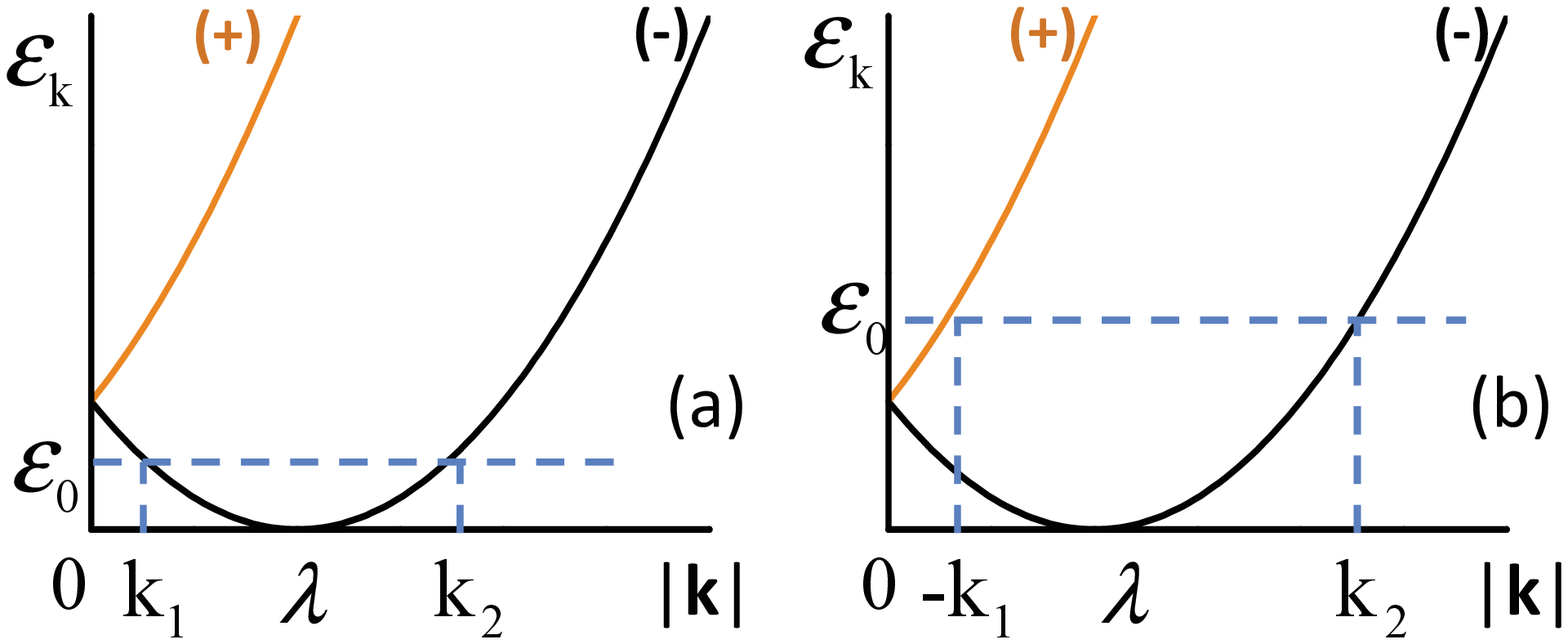}
\caption{(Color online) Single-particle spectrum,
$\epsilon^{(\pm)}_{\mathbf{k}}=(|\mathbf{k}|\pm\lambda)^2/(2m)$,
with isotropic SO coupling. For given energy
$\epsilon_0=k_0^2/(2m)$, two magnitudes of momentum are available as
$|k_1|$ and $k_2$, with $k_1=\lambda-k_0,\ k_2=\lambda+k_0$.
$k_0<\lambda$ in (a) and $k_0>\lambda$ in (b).} \label{fig_Ek}
\end{figure}

We start from the single-particle Hamiltonian of
spin$-1/2$($\uparrow, \downarrow$) fermions with isotropic SO
coupling, (we set the reduced Planck constant $\hbar=1$)
\begin{equation}
H_1=\frac{\mathbf{k}^2}{2m}+\frac{\lambda}{m}\mathbf{k}\cdot\boldsymbol\sigma+\frac{\lambda^2}{2m},
\end{equation}
where $\mathbf{k}=(k_x,k_y,k_z)$ and
$\boldsymbol\sigma=(\sigma_x,\sigma_y,\sigma_z)$ respectively denote
the momentum operator and Pauli spin operator; $\lambda$ is the
strength of SO coupling. The single-particle eigen-state has two
orthogonal branches as
\begin{eqnarray}
|\mathbf{k}^{(+)}\rangle&=&u^{(+)}_{\mathbf{k}}|\mathbf{k}_{\uparrow}\rangle
+u^{(-)}_{\mathbf{k}}
e^{i\phi_{\mathbf{k}}}|\mathbf{k}_{\downarrow}\rangle,\nonumber\\
|\mathbf{k}^{(-)}\rangle&=&-u^{(-)}_{\mathbf{k}}e^{-i\phi_{\mathbf{k}}}|\mathbf{k}_{\uparrow}\rangle
+u^{(+)}_{\mathbf{k}} |\mathbf{k}_{\downarrow}\rangle;
\end{eqnarray}
with $\phi_{\mathbf{k}}=arg(k_x+ik_y)$,
$u^{(\pm)}_{\mathbf{k}}=\sqrt{\frac{1}{2}\pm
\frac{k_z}{2|\mathbf{k}|}}$, and the corresponding eigen-energy
$\epsilon^{(\pm)}_{\mathbf{k}}=(|\mathbf{k}|\pm\lambda)^2/(2m)$ as
shown in Fig.\ref{fig_Ek}. Due to the isotropy of SO coupling, the
total angular momentum $\mathbf{j}=\mathbf{l}+\mathbf{s}\
(\mathbf{s}=\frac{1}{2}\boldsymbol\sigma)$
is conserved by $H_1$, 
giving the highest rotation symmetry among all types of SO
couplings.

The two-particle Hamiltonian can be written as
$H_2=H_{\mathbf{K}}+H_{\mathbf{k}}$, with $H_{\mathbf{K}}$ and
$H_{\mathbf{k}}$ respectively describing the center-of-mass motion
with total momentum $\mathbf{K}=\mathbf{k}_1+\mathbf{k}_2$ and
relative motion with momentum
$\mathbf{k}=(\mathbf{k}_2-\mathbf{k}_1)/2$,
\begin{eqnarray}
H_{\mathbf{K}}&=&\frac{\mathbf{K}^2}{4m}+\frac{\mathbf{K}}{4m}\cdot(I_1\otimes\boldsymbol\sigma_2+\boldsymbol\sigma_1\otimes
I_2),\\
H_{\mathbf{k}}&=&\frac{\mathbf{k}^2}{m}+\frac{\mathbf{k}}{m}\cdot(I_1\otimes\boldsymbol\sigma_2-\boldsymbol\sigma_1\otimes
I_2)+\frac{\lambda^2}{m}.\label{2-body-H}
\end{eqnarray}
With isotropic SO coupling, the total angular momentum for two
particles $\mathbf{J}=\mathbf{L}+\mathbf{S}$ is also conserved, with
$\mathbf{L}=\mathbf{l}_1+\mathbf{l}_2$,
$\mathbf{S}=\mathbf{s}_1+\mathbf{s}_2$ respectively the total
orbital angular momentum and total spin of particle $1$ and $2$.
Moreover, $\mathbf{L}$ can be decomposed as
$\mathbf{L}=\mathbf{L}_{\mathbf{r}}+\mathbf{L}_{\mathbf{R}}$, with
$\mathbf{L}_{\mathbf{r}}\ (\mathbf{L}_{\mathbf{R}})$ the angular
momentum for the relative motion
$\mathbf{r}=\mathbf{r}_2-\mathbf{r}_1$ (center-of-mass
$\mathbf{R}=(\mathbf{r}_1+\mathbf{r}_2)/2$). In view of the symmetry
of $H_2$, in this paper we consider the scattering problem in the
subspace of $\mathbf{K}=0(\mathbf{L}_\mathbf{R}=0)$ and
$\mathbf{J}=\mathbf{L}_\mathbf{r}+\mathbf{S}=0$. (The method
presented below can be generalized to the case of non-zero
$\mathbf{K}$ or $\mathbf{J}$).

For total $\mathbf{K}=0$, the scattered wavefunction only depends on
the relative coordinate $\mathbf{r}$, and is given by the
Lippmann-Schwinger equation\cite{book} as
\begin{eqnarray}
\langle \mathbf{r}|\Psi_{\mathbf{k}}\rangle&=&\langle
\mathbf{r}|\Psi_{\mathbf{k}}^{(0)}\rangle+\int d\mathbf{r}' \langle
\mathbf{r}|G(E)|\mathbf{r}' \rangle  \langle\mathbf{r}'|U
|\Psi_{\mathbf{k}}\rangle.\label{psi}
\end{eqnarray}
where $G(E)=\frac{1}{E-H_2+i\delta}$ is the two-particle green
function, $U$ is the interaction operator;
$|\Psi_{\mathbf{k}}^{(0)}\rangle$ is the incident two-particle state
with relative momentum $\mathbf{k}$, which can be either of the
following three states
\begin{eqnarray}
|\Phi_{\mathbf{k}}^{(--)}\rangle&=&|\mathbf{k}^{(-)},-\mathbf{k}^{(-)}\rangle
(-e^{i\phi_{\mathbf{k}}}) ,\\
|\Phi_{\mathbf{k}}^{(++)}\rangle&=&|\mathbf{k}^{(+)},-\mathbf{k}^{(+)}\rangle
(-e^{-i\phi_{\mathbf{k}}}) ,\\
|\Phi_{\mathbf{k}}^{(-+)}\rangle&=&|\mathbf{k}^{(-)},-\mathbf{k}^{(+)}\rangle
;
\end{eqnarray}
in coordinate space they are (we set the volume $V=1$ for
normalization)
\begin{widetext}
\begin{eqnarray}
\langle \mathbf{r}|\Phi_{\mathbf{k}}^{(--)}\rangle
&=&\frac{1}{\sqrt{2}}\Big\{-i\sin(\mathbf{k}\cdot
\mathbf{r})[-\frac{k_-}{k}|\uparrow\uparrow\rangle+\frac{k_+}{k}|\downarrow\downarrow\rangle+\frac{k_z}{k}(|\uparrow\downarrow\rangle+|\downarrow\uparrow\rangle)]
+\cos(\mathbf{k}\cdot
\mathbf{r})(|\uparrow\downarrow\rangle-|\downarrow\uparrow\rangle)\Big\},\label{2-body-psi_1}\\
\langle \mathbf{r}|\Phi_{\mathbf{k}}^{(++)}\rangle\rangle
&=&\frac{1}{\sqrt{2}}\Big\{i\sin(\mathbf{k}\cdot
\mathbf{r})[-\frac{k_-}{k}|\uparrow\uparrow\rangle+\frac{k_+}{k}|\downarrow\downarrow\rangle+\frac{k_z}{k}(|\uparrow\downarrow\rangle+|\downarrow\uparrow\rangle)]
+\cos(\mathbf{k}\cdot
\mathbf{r})(|\uparrow\downarrow\rangle-|\downarrow\uparrow\rangle)\Big\},\label{2-body-psi_2}\\
\langle \mathbf{r}|\Phi_{\mathbf{k}}^{(-+)}\rangle
&=&-\frac{1}{\sqrt{2}}i\sin(\mathbf{k}\cdot
\mathbf{r})\Big\{(1-\frac{k_z}{k})e^{-i\phi_{\mathbf{k}}}|\uparrow\uparrow\rangle+(1+\frac{k_z}{k})e^{i\phi_{\mathbf{k}}}|\downarrow\downarrow\rangle+
\frac{k_{\perp}}{k}(|\uparrow\downarrow\rangle+|\downarrow\uparrow\rangle)\Big\};\label{2-body-psi_3}
\end{eqnarray}
\end{widetext}
with $k=|\mathbf{k}|,\ k_{\perp}=\sqrt{k_x^2+k_y^2},\ k_{\pm}=k_x\pm
i k_y$.

Furthermore, the subspace of
$\mathbf{J}=\mathbf{L}_{\mathbf{r}}+\mathbf{S}=0$ can be spanned by
two orthogonal components as (labeled by $|L_r,S;m_L,m_s\rangle$)
\begin{eqnarray}
|J=0\rangle_0&=&|00;00\rangle,\\
|J=0\rangle_1&=&\frac{1}{\sqrt{3}}[|11;-1,1\rangle+|11;1,-1\rangle-|11;0,0\rangle].
\label{J0}
\end{eqnarray}
Here $|J=0\rangle_0$ is the spin-singlet combined with s-wave
orbital channel, and $|J=0\rangle_1$ is the spin-triplet combined
with p-wave orbital channel. Now any state projected to $J=0$
subspace can be written as
\begin{eqnarray}
\langle \mathbf{r}|\Psi_{\mathbf{k}} \rangle_{J=0}&=&\psi_0(r)
\langle\Omega_r|J=0\rangle_0+\psi_1(r)
\langle\Omega_r|J=0\rangle_1,\label{psi_P}
\end{eqnarray}
with bases
\begin{eqnarray}
\langle\Omega_r|J=0\rangle_0&=&Y_{00}(\Omega_r)\frac{|\uparrow\downarrow\rangle-|\downarrow\uparrow\rangle}{\sqrt{2}},\\
\langle\Omega_r|J=0\rangle_1&=&\frac{1}{\sqrt{3}}[Y_{1,-1}(\Omega_r)|\uparrow\uparrow\rangle+Y_{11}(\Omega_r)|\downarrow\downarrow\rangle \nonumber\\
&&-Y_{10}(\Omega_r)\frac{|\uparrow\downarrow\rangle+|\downarrow\uparrow\rangle}{\sqrt{2}}],
\end{eqnarray}
and wavefunctions
\begin{eqnarray}
\psi_0(r)&=&\int d\Omega_r \ {_0}\langle J=0|\Omega_r
\rangle \langle \mathbf{r}|\Psi_{k}\rangle; \\
\psi_1(r)&=&\int d\Omega_r \ {_1}\langle J=0|\Omega_r \rangle
\langle \mathbf{r}|\Psi_k\rangle. \label{psi0_P}
\end{eqnarray}
Here $\Omega_r$ denotes the azimuthal angle of relative coordinate
$\mathbf{r}$, and $Y_{lm}$ the spherical harmonics with azimuthal
quantum numbers $(l,m)$.  After projected, the eigen-states of
$H_2$, i.e.,
Eqs.(\ref{2-body-psi_1},\ref{2-body-psi_2},\ref{2-body-psi_3}), are
given by
\begin{eqnarray}
\langle \mathbf{r}|\Phi_{\mathbf{k}}^{(--)}
\rangle_{J=0}&=&\sqrt{4\pi}[j_0(kr)\langle\Omega_r|J=0\rangle_0+
\nonumber\\
&&\ \ \ \ \ \ \ i
j_1(kr)\langle\Omega_r|J=0\rangle_1 ], \label{psi0_1}\\
\langle \mathbf{r}|\Phi_{\mathbf{k}}^{(++)}
\rangle_{J=0}&=&\sqrt{4\pi}[j_0(kr)\langle\Omega_r|J=0\rangle_0- \nonumber\\
&&\ \ \ \ \ \ \ i
j_1(kr)\langle\Omega_r|J=0\rangle_1 ], \label{psi0_2}\\
\langle \mathbf{r}|\Phi_{\mathbf{k}}^{(-+)} \rangle_{J=0}&=&0.
\label{psi0_3}
\end{eqnarray}
with $j_l(x) (l=0,1)$ the spherical Bessel function of $l-$th order.
Particularly, Eq.\ref{psi0_3} shows that $(-+)$ channel is not
involved in the subspace of $J=0$.

Due to the single-particle spectrum modified by isotropic SO
coupling (see Fig.\ref{fig_Ek}), the incident state of two particles
with energy $E=k^2/m$ can be an arbitrary combination of plane-waves
with two different magnitudes of momenta, $|\mathbf{k}_2|=\lambda+k$
and $|\mathbf{k}_1|=|\lambda-k|$. For $k<\lambda$,
\begin{eqnarray}
\langle\mathbf{r}|\Psi_{\mathbf{k}}^{(0)}\rangle_{J=0}&=&\alpha\langle\mathbf{r}|\Phi_{\mathbf{k_2}}^{(--)}\rangle_{J=0}
+\beta\langle\mathbf{r}|\Phi_{\mathbf{k_1}}^{(--)}\rangle_{J=0};
\end{eqnarray}
and for $k>\lambda$,
\begin{eqnarray}
\langle\mathbf{r}|\Psi_{\mathbf{k}}^{(0)}\rangle_{J=0}&=&\alpha\langle\mathbf{r}|\Phi_{\mathbf{k_2}}^{(--)}\rangle_{J=0}
+\beta\langle\mathbf{r}|\Phi_{\mathbf{k_1}}^{(++)}\rangle_{J=0},
\end{eqnarray}
which both result in ($k_2\equiv\lambda+k,\ k_1\equiv\lambda-k$)
\begin{widetext}
\begin{eqnarray}
\langle\mathbf{r}|\Psi_{\mathbf{k}}^{(0)}\rangle_{J=0}&=&\sqrt{4\pi}\{[\alpha
j_0(k_2r)+\beta j_0(k_1r)]\langle\Omega_r|J=0\rangle_0+i[\alpha
j_1(k_2r)+\beta j_1(k_1r)]\langle\Omega_r|J=0\rangle_1\}.
\label{psi0_P}
\end{eqnarray}

In view of the property of $H_2$, we also project the interaction
$U$ (with range $r_0$) to $J=0$ subspace as
\begin{eqnarray}
\langle \mathbf{r}|U|\Psi_{\mathbf{k}} \rangle_{J=0}&=&\sqrt{4\pi}
[F_0(r) \langle\Omega_r|J=0\rangle_0+F_1(r)
\langle\Omega_r|J=0\rangle_1], \ \ \ (r<r_0)\label{U_P}
\end{eqnarray}
here $F_0,\ F_1$ denote the scattering amplitude in
s-wave($L_{\mathbf{r}}=0$) and p-wave($L_{\mathbf{r}}=1$) channel.
The Green function in Eq.\ref{psi} is calculated by inserting a
complete set of intermediate states (Eq.\ref{psi0_1} and
\ref{psi0_2}),
\begin{eqnarray}
\langle
\mathbf{r}|G|\mathbf{r}'\rangle_{J=0}&=&\frac{1}{2}\sum_{\mathbf{k}}\big\{
\frac{\langle \mathbf{r}|\Phi_{\mathbf{k}}^{(--)}\rangle \langle
\Phi_{\mathbf{k}}^{(--)}|\mathbf{r'}\rangle}{E-2\epsilon_{\mathbf{k}}^{(-)}+i\delta}
+ \frac{\langle \mathbf{r}|\Phi_{\mathbf{k}}^{(++)}\rangle
\langle\Phi_{\mathbf{k}}^{(++)}|\mathbf{r'}
\rangle}{E-2\epsilon_{\mathbf{k}}^{(+)}+i\delta}\big\}_{J=0}.\label{G_P}
\end{eqnarray}
Here the prefactor $1/2$ is to eliminate the double counting of
inserted states.

Combining Eqs.(\ref{psi}, \ref{psi0_P}, \ref{U_P}, \ref{G_P}), we
obtain the closed form of scattered wavefunction (for $r>r_0$) in
each partial-wave channel (see Eq.\ref{psi_P}) as
\begin{eqnarray}
\psi_0/\sqrt{4\pi}&=&\alpha j_0(k_2r)+C_{k_2}[n_0(k_2r)-i
j_0(k_2r)]+ \beta
j_0(k_1r)+C_{k_1}[n_0(k_1r)+i j_0(k_1r)],\nonumber\\
\psi_1/(i\sqrt{4\pi})&=&\alpha j_1(k_2r)+C_{k_2}[n_1(k_2r)-i
j_1(k_2r)]+ \beta j_1(k_1r)+C_{k_1}[n_1(k_1r)+i j_1(k_1r)]
\label{psi_01_super}
\end{eqnarray}
\end{widetext}
where ($q=k_1$ or $k_2$)
\begin{eqnarray}
C_{q}&=&\frac{q^2}{2(q-\lambda)}(f_0(q)-if_1(q)),\label{C}\\
f_0(q)&=&m\int_0^{r_0} dr r^2 F_0(r)j_0(qr),\label{f_0}\\
f_1(q)&=&m\int_0^{r_0} dr r^2 F_1(r)j_1(qr),\label{f_1}
\end{eqnarray}
and $n_l(x)\ (l=0,1)$ the spherical Neumann function of $l-$th
order.

We further simplify the complex wavefunction (\ref{psi_01_super}) by
employing the time-reversal symmetry, i.e., $[H_2,T]=0$ where $T$ is
the time-reversal operator. Therefore we choose the wavefunction to
be the eigen-state for both $H_2$ and $T$. Noting that
$T\langle\Omega_r|J=0\rangle_0=\langle\Omega_r|J=0\rangle_0$,
$T\langle\Omega_r|J=0\rangle_1=-\langle\Omega_r|J=0\rangle_1$, the
only way to achieve $T\Psi=e^{i\theta}\Psi$ is to assume
\begin{eqnarray}
C_{k_2}&=&-\alpha \sin\delta e^{i\delta},\label{Ck2}\\
C_{k_1}&=&\beta \sin\delta e^{i\delta},\label{Ck1}
\end{eqnarray}
with $\delta=\theta/2$. Then up to a prefactor
$\sqrt{4\pi}\cos\delta e^{i\delta}$, Eq.\ref{psi_01_super} is
reduced to

\begin{eqnarray}
\psi_0&=&\alpha [j_0(k_2r)-\tan\delta n_0(k_2r)]+\nonumber\\
&& \ \ \ \ \ \beta[j_0(k_1r)+\tan\delta
n_0(k_1r)],\nonumber\\
\psi_1/i&=&\alpha [j_1(k_2r)-\tan\delta n_1(k_2r)]+\nonumber\\
&& \ \ \ \ \ \beta[j_1(k_1r)+\tan\delta n_1(k_1r)].
\label{psi_01_super2}
\end{eqnarray}

To this end we have obtained the exact form of scattered
wavefunction for a given short-range potential $U(\mathbf{r})$
defined in Eq.\ref{U_P}. Eq.\ref{psi_01_super2} reveals a unique
scattering property in the presence of isotropic SO coupling, i.e.,
the wavefunction in each partial-wave channel is characterized by
two different momenta(see also Fig.\ref{fig_Ek}) with opposite phase
shifts. Note that without SO coupling, $\lambda=0$, $k_2=-k_1=k$,
Eq.\ref{psi_01_super2} reduces to the standard form of s-wave and
p-wave scattered wavefunctions in free space.

The scattered wavefunction (Eq.\ref{psi_01_super2}) has the
following asymptotic behaviors at long-range and short-range of
inter-particle distances. As $kr\rightarrow\infty$, the long-range
behavior is (up to a prefactor $\sqrt{4\pi}e^{i\delta}$)
\begin{eqnarray}
\psi_0&=& \alpha \frac{\sin(k_2r+\delta)}{k_2r}+
\beta\frac{\sin(k_1r-\delta)}{k_1r},\label{psi_0_long} \\
\psi_1/i&=& \alpha \frac{\sin(k_2r-\pi/2+\delta)}{k_2r}+
\beta\frac{\sin(k_1r-\pi/2-\delta)}{k_1r}. \label{psi_1_long}
\end{eqnarray}
At short-range $r_0<r\ll 1/k$, we have (up to a prefactor
$\sqrt{4\pi}\cos\delta e^{i\delta}$)
\begin{eqnarray}
\psi_0&=&\alpha
+\beta+(\frac{\alpha}{k_2}-\frac{\beta}{k_1})\frac{\tan\delta}{r}, \label{short_psi0} \\
\psi_1/i&=&\frac{\alpha k_2+\beta
k_1}{3}r+(\frac{\alpha}{k_2^2}-\frac{\beta}{k_1^2})\frac{\tan\delta}{r^2}.
\label{short_psi1}
\end{eqnarray}

For simplicity, we consider the limit of zero-range potential, i.e.,
assuming $F_i(r)\equiv\frac{\delta(r)}{4\pi
r^2}\overline{F}_i(r\rightarrow0)\ (i=0,1)$ in Eq.\ref{U_P}. Further
according to Eqs.(\ref{f_0},\ref{f_1}) we introduce
\begin{equation}
\overline{f}_0=\frac{m}{4\pi}\overline{F}_0(r\rightarrow0),\ \ \
\overline{f}_1=\frac{m}{4\pi}\frac{r\overline{F}_1(r\rightarrow0)}{3}|_{r\rightarrow0},
\label{f_01}
\end{equation}
which gives $f_0(q)=\overline{f}_0,\ f_1(q)=q\overline{f}_1$
$(q=k_2$ or $k_1)$. Eqs.(\ref{C},\ref{Ck2},\ref{Ck1}) then relate
$\overline{f}_0$ and $\overline{f}_1$ to $\alpha,\beta,\delta$ as
\begin{eqnarray}
\overline{f}_0&=&\sin\delta e^{i\delta}(\frac{\alpha
k_1}{k_2^2}-\frac{\beta k_2}{k_1^2}),\\
 i\overline{f}_1&=&\sin\delta
e^{i\delta}(\frac{\alpha}{k_2^2}-\frac{\beta}{k_1^2}).\label{f10}
\end{eqnarray}
Thus the short-range behavior(Eqs.\ref{short_psi0},\ref{short_psi1})
can be expressed in terms of $\overline{f}_0, \overline{f}_1$ as (up
to a prefactor $4\pi\cot\delta$)
\begin{eqnarray}
\psi_0&=&\frac{i\overline{f}_1(k_1^3+k_2^3)-\overline{f}_0(k_1^2+k_2^2)}{k_2-k_1}+\nonumber\\
&&\ \ \ \ \ \ \ \ (i\overline{f}_1(k_1+k_2) -\overline{f}_0)\frac{\tan\delta}{r};\label{short_psi0_f} \\
\psi_1/i&=&
\frac{i\overline{f}_1(k_1^4+k_2^4)-\overline{f}_0(k_1^3+k_2^3)}{3(k_2-k_1)}
r +i\overline{f}_1\frac{\tan\delta}{r^2}. \label{short_psi1_f}
\end{eqnarray}
These results show that with SO coupling, the short-range
parametrization of the wavefunction in each partial-wave channel
will additionally depend on scattering amplitude of another
partial-wave channel. This directly reflects the spin-mediated mixed
scattering between different orbital (partial-wave) channels, as is
one of the most dramatic features of SO coupled system.

At the end of this section, we study the bound state solution with
energy $E=-\kappa^2/m<0$. The bound state is given by the poles of
scattering amplitudes $(\bar{f}_0,\bar{f}_1\rightarrow\infty)$,
which corresponds to the following transformation from the
scattering state\cite{Stock05}
\begin{equation}
k\rightarrow i\kappa, \ \ \ \delta\rightarrow -i\infty.\label{bs}
\end{equation}
Using Eq.\ref{bs}, the bound state wavefunction can be deduced from
Eq.\ref{psi_01_super2}; its long-range and short-range behaviors can
be deduced from Eqs.(\ref{psi_0_long},\ref{psi_1_long}) and
Eqs.(\ref{short_psi0},\ref{short_psi1}) respectively.

\section{Pseudo-potential model in individual partial-wave channel}

The pseudo-potential model formulated in a given partial-wave
channel is based on two assumptions. First, the interaction only
acts on this particular channel. Second, the short-range behavior of
wavefunction in this channel is still determined by the same
scattering parameter as that in the absence of SO coupling. The
second assumption is based on a general belief as follows. If the
range of interacting potential ($r_0$) is much shorter than any
length scale in the system, as inter-particle distance approaches
$r\rightarrow r_0^+$, all other potentials are negligible in this
limit and the asymptotic behavior of two-body wavefunction is
unchanged. The validity of pseudo-potentials has been verified in
trapped systems in Ref.\cite{Stock05, E-dependence, Yip}. In the
following we reduce the exact solutions obtained in Section II to
the framework of s-wave and p-save pseudo-potential models.

\subsection{s-wave pseudo-potential}

The s-wave pseudo-potential corresponds to assuming
$\overline{F}_1=0, \overline{f}_1=0$; by mapping the short-range
behavior of $\psi_0$(Eq.\ref{short_psi0_f}) to $1/r-1/a_s$ with
$a_s$ the s-wave scattering length in free space, we obtain the
phase shift as
\begin{equation}
\tan\delta=-a_s \frac{\lambda^2+k^2}{k}.\label{phase_s-pseudo}
\end{equation}

For scattering state, at low energies, $\tan\delta=-a_s\lambda^2/k$
giving the effective 1D coupling $g_{1D}=2a_s\lambda^2/m$, which is
supported by the modified low-energy density of state(DOS) by
isotropic SO couplings (see also Ref.\cite{2-body}); at high
energies, Eq.\ref{phase_s-pseudo} reduces to $\tan\delta=-ka_s$ as
in 3D free space.

The equation for bound state solution is obtained from
Eq.\ref{phase_s-pseudo} via transformations as Eq.\ref{bs},
\begin{equation}
-\frac{1}{a_s}\kappa=\lambda^2-\kappa^2,\label{bound_s-pseudo}
\end{equation}
which reproduces the result obtained by s-wave T-matrix
approach\cite{2-body}. Eq.\ref{bound_s-pseudo} results in a bound
state solution for arbitrarily weak interaction, which is a direct
consequence of the effective 1D DOS at low energies.

\subsection{p-wave pseudo-potential}

The p-wave pseudo-potential corresponds to $\overline{F}_0=0,
\overline{f}_0=0$, and $\delta$ is determined by mapping the
short-range behavior of $\psi_1$(Eq.\ref{short_psi1_f}) to
$r/3-v_p/r^2$, with $v_p$ the p-wave scattering volume in free
space. We obtain
\begin{equation}
\tan\delta=-v_p \frac{\lambda^4+6\lambda^2k^2
+k^4}{k}.\label{phase_p-pseudo}
\end{equation}
Without SO coupling ($\lambda=0$), it reproduces the original free
space result as $\tan\delta=-v_p k^3$.

For scattering state at low energies, $\tan\delta=-v_p
\frac{\lambda^4}{k}$ again giving $\delta(k=0)=\pi/2$; at high
energies, it recovers the free space result.

For bound state, by transformation as Eq.\ref{bs} we obtain from
Eq.\ref{phase_p-pseudo} that
\begin{equation}
-\frac{1}{v_p}\kappa= \lambda^4-6\lambda^2\kappa^2
+\kappa^4.\label{bound_p-pseudo}
\end{equation}
We see that for arbitrarily weak p-wave interaction
$v_p\rightarrow0^-$, Eq.\ref{bound_p-pseudo} gives a shallow bound
state as $\kappa=-v_p \lambda^4$.

\section{Scattering under a square-well potential and Validity of
pseudo-potentials}

In this section we present the scattering state and bound state
solutions under a square-well interaction potential. By comparing
these solutions with those from individual s-wave and p-wave
pseudo-potential model, we shall address the validity of
pseudo-potentials in the presence of isotropic SO coupling. In
Appendix A we show more details about partial-wave scattering under
the square-well potential without SO coupling, and in Appendix B we
derive the equations for two-body solutions with isotropic SO
coupling.

\subsection{Results}

We consider a square-well potential with depth $V_0(<0)$ at
inter-particle distance $r<r_0$ and with depth zero otherwise. The
interaction strength is uniquely characterized by a dimensionless
parameter as $qr_0$, with $q=\sqrt{-mV_0}$. Without SO coupling,
Eq.\ref{delta_SW} shows that by increasing $qr_0$, a sequence of
s-wave resonances(with phase shift $\delta_s=\pi/2$) occur at
$qr_0/\pi=n+1/2$ and p-wave resonances($\delta_p=\pi/2$) at
$qr_0/\pi=n+1$ ($n=0,1,2...$). A bound state emerges whenever across
a scattering resonance.

Next we solve the two-body problem in the presence of isotropic SO
coupling. Based on exact solutions in section II, the wavefunctions
inside the potential ($r<r_0$) in orbital s-wave and p-wave channels
are
\begin{eqnarray}
\psi_0&=&j_0(q_2r)+ t j_0(q_1r),\nonumber\\
\psi_1/i&=&j_1(q_2r)+ t j_1(q_1r); \label{small_r}
\end{eqnarray}
with $q_2=\lambda+\sqrt{m(E-V_0)},\ q_1=\lambda-\sqrt{m(E-V_0)}$.
Outside the potential ($r>r_0$), the wavefunctions are given by
Eq.\ref{psi_01_super2} for the scattering state($E=k^2/m>0$), or by
the transformed form (through Eq.\ref{bs}) for bound state
($E=-\kappa^2/m<0$).

Using the continuity properties of $\psi_0$, $\psi_1$ and their
first-order derivatives at the boundary $r=r_0$, one can solve all
the unknown parameters $\{t,\ \alpha,\ \beta,\ \delta\}$ for the
scattering state and $\{t,\alpha,\beta,\kappa\}$ for the bound
state. In Appendix \ref{SO_SW} we present the equations for these
solutions. Next we show numerical results for the scattering state
and bound state in turn.

\subsubsection{Scattering state}

\begin{figure}[hbtp]
\includegraphics[height=5.2cm,width=8.8cm]{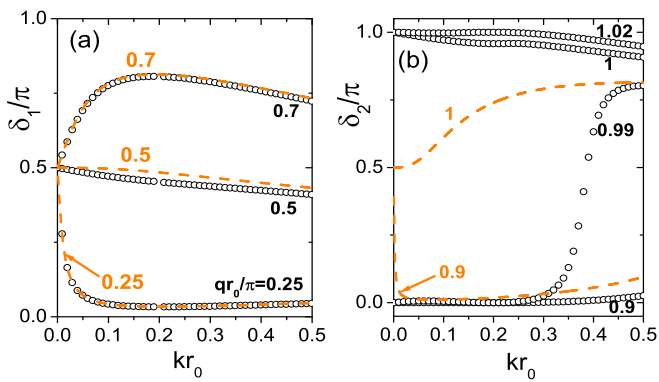}
\caption{(Color online) Phase shifts of scattering states in a
square-well potential with isotropic SO coupling strength $\lambda
r_0=0.2$. Exact solutions are shown (by black circles) in comparison
with results from pseudo-potential models with effective-range
corrections(see Eq.\ref{eff-range}) (by orange dashed lines).
(a)$\delta_1$ near s-wave resonance with (from bottom to top)
$qr_0/\pi=0.25(a_s<0), 0.5(a_s=\infty), 0.7(a_s>0)$. (b)$\delta_2$
near p-wave resonance with (from bottom to top)
$qr_0/\pi=0.9(a_p<0), 0.99(a_p<0), 1(a_p=\infty), 1.02(a_p>0)$. }
\label{fig_upper}
\end{figure}

For given energy $E=k^2/m>0$, we obtain two phase shift solutions,
$\delta_1$ with $\bar{f}_0\gg \bar{f}_1$ and $\delta_2$ with
$\bar{f}_1\gg \bar{f}_0$, analogous to s-wave and p-wave phase
shifts without SO coupling. For fixed SO coupling $\lambda r_0=0.2$,
we show in Fig.\ref{fig_upper}(a) the solution of $\delta_1$ near
the first s-wave resonance and in Fig.\ref{fig_upper}(b) the
solution of $\delta_2$ near the first p-wave resonance.
Independently, we obtain $\delta_1$ from Eq.\ref{phase_s-pseudo}
using s-wave scattering length($a_s$) with effective-range
corrections (see Eq.\ref{eff-range}, $l=0$), and $\delta_2$ from
Eq.\ref{phase_p-pseudo} using p-wave scattering length($a_p\equiv
v_p/r_0^2$) with effective-range corrections (Eq.\ref{eff-range},
$l=1$). In Fig.\ref{fig_upper}, these results are shown (by orange
dashed lines) to compare with exact solutions (black circles).

\begin{figure}[hbtp]
\includegraphics[height=5cm,width=8.5cm]{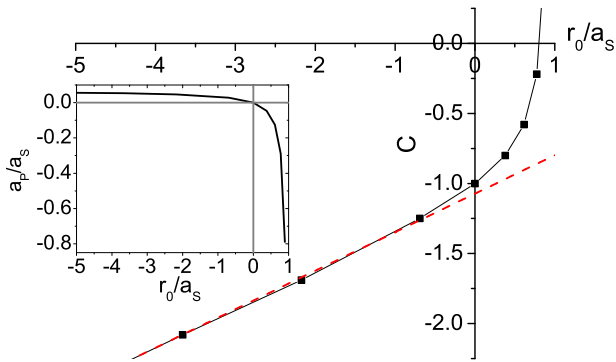}
\caption{(Color online) $C$ in Eq.\ref{equ_C} as functions of
$r_0/a_s$ near and across the first s-wave resonance ($qr_0<\pi$) in
the square-well potential. The red dashed line is the linear fit in
the weak interaction limit as $C=-1.073+0.275r_0/a_s$. Inset shows
the ratio of p-wave to s-wave scattering length, where the gray
(light) lines denote the s-wave resonance.} \label{fig_C}
\end{figure}

For the solution $\delta_1$ near s-wave resonance,
Fig.\ref{fig_upper}(a) shows that it can be approximately fit by
s-wave model within $kr_0<0.5$. Particularly at $k=0$,
$\delta_1=\pi/2$ is consistent with the s-wave prediction
(Eq.\ref{phase_s-pseudo}) due to the 1D feature of the low-energy
DOS. However, there is still a small deviation between these two
solutions at finite $k$, due to the interplay between SO coupling,
p-wave contribution and the finite potential range. To investigate
these effects in detail, we further study the modified effective
scattering length $a_{eff}$ in s-wave channel, which is defined by
$\psi_0(r)\rightarrow 1/r-1/a_{eff}(k)$ at $r_0<r\ll 1/k$.
Practically $a_{eff}(k)$ can be extracted from the asymptotic
wavefunction (\ref{short_psi0_f}) by diagonalizing
Eq.\ref{matrix_SW} in Appendix B. Fig.\ref{fig_short-range}(a) shows
how $a_{eff}(k)$ evolves with $k$ for each given SO coupling, which
can also be expressed in the form of effective-range correction,
\begin{equation}
\frac{1}{a_{eff}(k)}=\frac{1}{a_{eff}}-\frac{1}{2}r_{eff}k^2.
\end{equation}
One can see that with increased SO couplings, the effective range
$r_{eff}$ almost stay unchanged while $1/a_{eff}$ become smaller
indicating weaker interactions. The deviations of $1/a_{eff}$ from
$1/a_s$ directly manifest the effect of SO coupling and mixed
scattering of s-wave channel with p-wave channel. Moreover the
mixing can also been seen from the additional dependence of $\psi_0$
on the p-wave scattering amplitude in Eq.\ref{short_psi0_f}. In
Fig.\ref{fig_short-range}(b) we show the zero-energy value
$1/a_{eff}$ as a function of $\lambda r_0$ for several different
potential depths. At $\lambda r_0\ll 1$, $1/a_{eff}$ can be well fit
by
\begin{equation}
\frac{r_0}{a_{eff}}=\frac{r_0}{a_s}+C(\lambda r_0)^2,\label{equ_C}
\end{equation}
where the dimensionless parameter $C$ only depends on the properties
of the potential, or the actual interaction strengths in s-wave and
p-wave channels. In Fig.\ref{fig_C}, $C$ is shown as a function of
$r_0/a_s$ (together with $a_p/a_s$) near the first s-wave resonance.
In the weak interaction limit, $|V_0|\rightarrow0$ and
$a_s,a_p\rightarrow 0^-$, $C$ change linearly with $r_0/a_s$,
indicating $a_{eff}-a_s\propto-(\lambda r_0)^2$ in this limit. For
the typical parameter regime in the present
experiment\cite{expe_soc}, $\lambda$ is determined by the wavevector
of the laser which is much smaller than the cutoff momentum of
realistic potential. In this case, the condition $\lambda r_0\ll 1$
gives negligible correction to $a_{eff}$ near s-wave resonances.

\begin{figure}[hbtp]
\includegraphics[height=5.1cm,width=9cm]{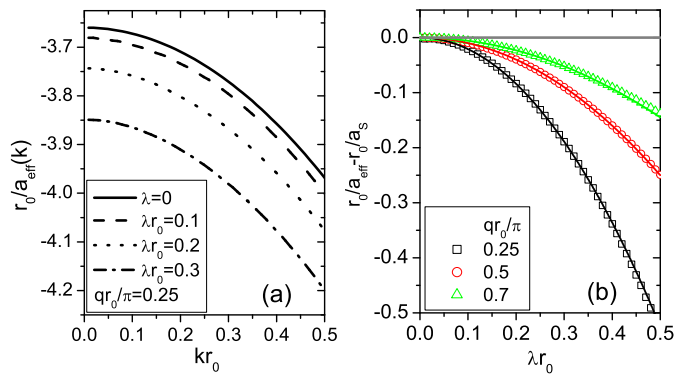}
\caption{(Color online) (a)Inverse of effective scattering lengths
($a_{eff}(k)$) as functions of $k$ for different SO couplings
$\lambda r_0=0, 0.1, 0.2, 0.3$(from top to bottom). The potential
depth is $qr_0/\pi=0.25$. (b) Inverse of zero-energy scattering
lengths ($a_{eff}$) as functions of $\lambda$ at different
$qr_0/\pi=0.25, 0.5, 0.7$(from bottom to top). The lines are fits to
Eq.\ref{equ_C} with $C=-2.10, -1.00, -0.58$(from bottom to top). }
\label{fig_short-range}
\end{figure}

For the solution $\delta_2$ near p-wave resonance, however, it
behaves qualitatively different from that obtained entirely in the
framework of p-wave pseudo-potential model, as showed by
Fig.\ref{fig_upper}(b). Obviously, the exact solution shows the
initial value $\delta_2(k=0)=0$ or $\pi$, depending on whether or
not there is a two-body bound state(see next section); while the
p-wave model always predicts $\delta_2(k=0)=\pi/2$ according to
Eq.\ref{phase_p-pseudo}. We have checked that in the limit of
$\lambda r_0\ll1$, the exact solution of $\delta_2$ at $k\ll1/r_0$
essentially follows the free space result (given by
$\tan\delta_2=-v_p k^3$) with $\delta_2\sim0$ or $\pi$; while the
p-wave model gives a narrow momentum window as $0<k<\lambda$ when
$\delta_2(k)$ evolves from $\pi/2$ to the exact result. This
dramatic difference indicates that even near the p-wave resonance,
the p-wave pseudo-potential alone can not be applied to the fermion
system if $\lambda$ is larger or comparable to the Fermi momentum.
We shall analyze the reason for the breakdown of p-wave model to
scattering state solutions in the discussion section.

\subsubsection{Bound state}

\begin{figure}[hbtp]
\includegraphics[height=6.3cm,width=7.5cm]{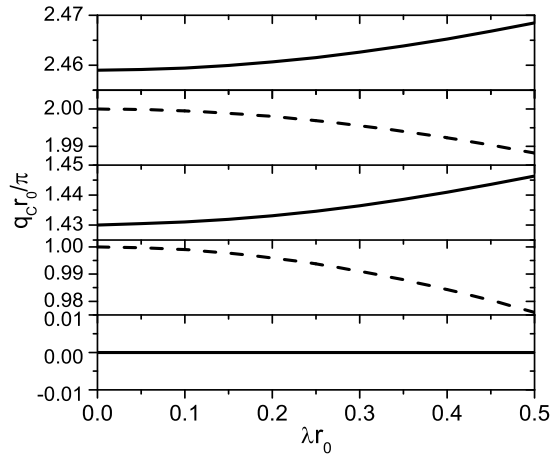}
\caption{ Critical potential depth $(q_cr_0/\pi)$ for the emergence
of each new bound state as a function of isotropic SO coupling
strength. When $\lambda r_0\rightarrow 0$, the solid lines approach
$q_cr_0/\pi=0, 1.430,2.459,...$ with $a_s=0$; the dashed lines
approach $q_cr_0/\pi=1, 2,...$ with $a_p=\infty$ (See text). }
\label{fig_qc}
\end{figure}

The bound state solution $E=-\kappa^2/m<0$ is given by the
transformed matrix equation $|A_b|=0$(see Appendix B). By setting
$\kappa=0$ in the matrix equation we determine the critical
potential depth $|V_0|_c$, which is responsible for the emergence of
a new bound state, by
\begin{equation}
j_0(q_2r_0)j_1(q_1r_0)=j_0(q_1r_0)j_1(q_2r_0),\label{crit_V}
\end{equation}
with $q_2=\lambda+q_c, q_1=\lambda-q_c$ and $q_c=\sqrt{m|V_0|_c}$.
The solution of $q_c$ is shown in Fig.\ref{fig_qc}. As $\lambda$
approaches zero, one branch of solution(solid line) is given by
$j_1(q_cr_0)=0$ or $a_s=0$; the other branch(dashed line) is given
by $j_0(q_cr_0)=0$ or $a_p=\infty$. For the first branch, when
increasing $\lambda$ the lowest solution will stay at $q_cr_0=0$ or
$a_s=0^-$, while the other solutions increase resulting in deeper
potential depths. For the second branch, when increasing $\lambda$
all solutions of $q_c$ will decrease, implying that weaker
interaction is required to support the new bound state near p-wave
resonance. In all, we see that only the lowest solution of the first
branch is consistent with the prediction from s-wave
pseudo-potential model (see Eq.\ref{bound_s-pseudo}), but none of
the other solutions. The discrepancies here are attributed to the
mixed scattering between s-wave and p-wave channels induced by the
isotropic SO coupling.

\begin{figure}[hbtp]
\includegraphics[height=8cm,width=7.5cm]{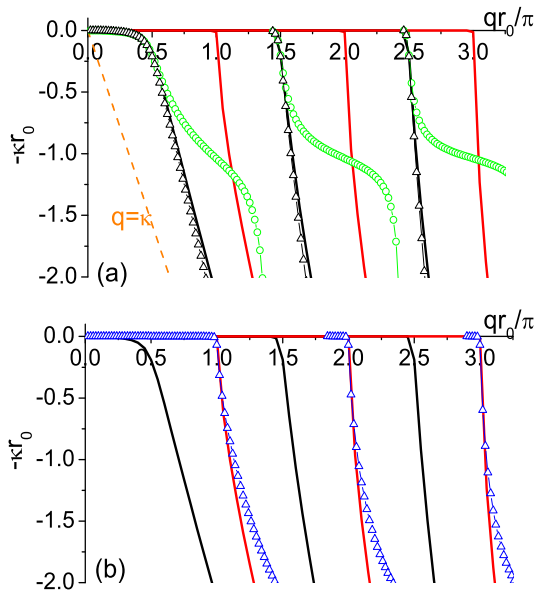}
\caption{(Color online) Bound state solutions (solid lines) in the
square-well potential as functions of the potential depths, in
comparison with results from s-wave(a) and p-wave(b)
pseudo-potential models. The SO coupling is $\lambda r_0=0.2$. In
(a), we have used s-wave scattering length without(green circles) or
with (black triangles) energy-dependence. In (b), we use the
energy-dependent p-wave scattering length(blue triangles).}
\label{fig_Eb}
\end{figure}

As shown in Fig.\ref{fig_Eb} with fixed $\lambda r_0=0.2$, a
sequence of bound states will develop when the potential depths
increase above critical $|V_0|_c$. For comparison, we also present
the results from s-wave and p-wave pseudo-potential models, using
the scattering length with or without energy-dependence. (For the
bound state, the energy-dependent scattering length is determined
from Eqs.(\ref{al},\ref{delta_SW}) but with $k$ replaced by
$i\kappa$. \cite{Stock05})

\begin{figure}[hbtp]
\includegraphics[height=8cm,width=8.8cm]{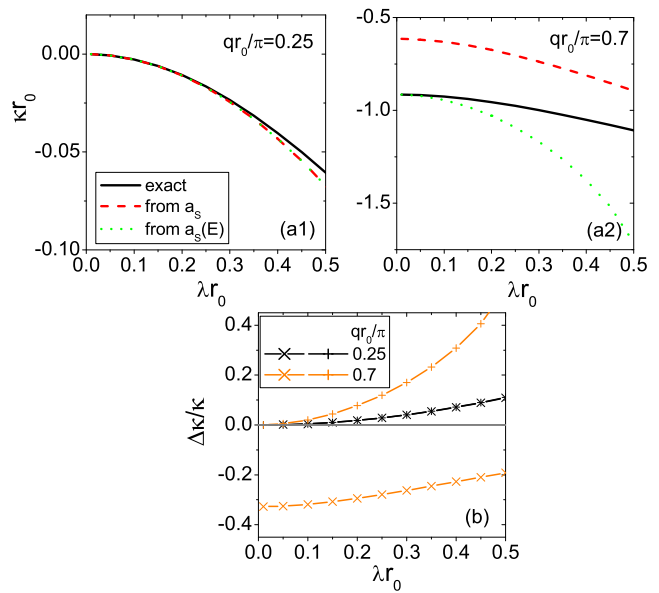}
\caption{(Color online) (a1,a2) Bound state solutions in a
square-well potential as functions of isotropic SO coupling
strengths at potential depths $qr_0/\pi=0.25[(a1),\ a_s<0]$ and
$0.7[(a2),\ a_s>0]$, in comparison with results using s-wave
scattering length without (red dashed line) or with (green dot line)
energy-dependence. (b) Relative deviations between exact solutions
and results from s-wave pseudo-potential model, at the same
potential depths as in (a1,a2). For s-wave pseudo-potential, we use
s-wave scattering length without ($'\times'$) or with ($'+'$)
energy-dependence. } \label{fig_Eb-corr}
\end{figure}

Fig.\ref{fig_Eb}(a) shows that the s-wave model using s-wave
scattering length without($a_s$) or with($a_s(E)$) energy-dependence
both give good approximations to low-energy solutions near s-wave
resonance, but deviate a lot from exact solutions for deep bound
states. In general, we find that using $a_s(E)$ provides more
accurate results than using $a_s$ in a large energy-range;
particularly, in the limit of zero SO coupling, using $a_s(E)$ will
give the exact bound state solutions\cite{Stock05}. For fixed
potential depth, the deviation of s-wave results from exact
solutions increases with the SO coupling strength, as shown in
Fig.\ref{fig_Eb-corr}(a1,a2). Moreover, Fig.\ref{fig_Eb}(a) and
Fig.\ref{fig_Eb-corr} show that the s-wave models using $a_s(E)$
always predict deeper bound states than real solutions, which is
consistent with Eq.\ref{equ_C} and also Fig.\ref{fig_short-range}
that the presence of SO coupling reduces the effective interaction
parameter for low-energy states.

\begin{figure}[hbtp]
\includegraphics[height=4.2cm,width=6.5cm]{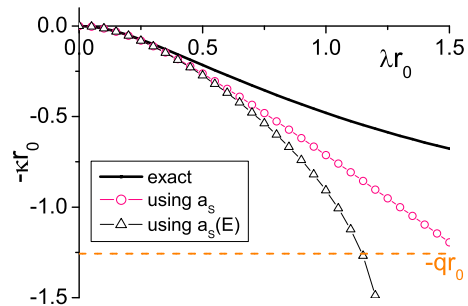}
\caption{(Color online) Bound state solutions as functions of
isotropic SO coupling strengths for given potential depth
$qr_0/\pi=0.4$, in comparison with results using s-wave scattering
length without(pink circles) or with(black triangles)
energy-dependence. The orange dashed line denotes the lower bound as
$\kappa=q$. } \label{fig_low-bound}
\end{figure}

In Fig.\ref{fig_Eb-corr}(b), we further plot the relative
deviations, $\Delta\kappa/\kappa$, as functions of SO coupling
strengths at different potential depths. It shows that
$\Delta\kappa/\kappa$ increases more rapidly for deep bound states
than that for shallow ones. As also mentioned in the introduction,
the s-wave model(even using the energy-dependent $a_s(E)$) is quite
questionable when applied to deep molecules. As shown in
Fig.\ref{fig_low-bound}, the energy of the bound state is always
lower bounded by the potential depth $V_0$, i.e., $\kappa<q$.
However, the s-wave model will produce unphysically deep molecules
with $\kappa>q$. In this case, the s-wave model alone will not work
and one must take into account the effect mixed scattering with
p-wave channel due to SO couplings.


In Fig.\ref{fig_Eb}(b) we show the comparison with results from
p-wave pseudo-potential model. According to the p-wave model (see
Section IIIB), the bound state exists for an arbitrarily weak
interaction in the presence of isotropic SO coupling. This is
qualitatively different from the exact solution under the
square-well potential, where each emergence of a new bound state
requires a potential depth beyond the critical value (as shown by
red lines in Fig.\ref{fig_qc}). In the limit of $\lambda r_0\ll1$,
the critical depths continuously approach $q_cr_0/\pi=1,2,...$ as in
free space. The breakdown of p-wave model to bound state solutions
will be discussed in the following section.

\subsection{Discussion}

Through the last subsection, we have shown that the SO coupling has
different effects on the validity of pseudo-potentials in s-wave and
p-wave channels. In the limit of $\lambda r_0\ll 1$, the s-wave
pseudo-potential model provides good approximations to the
low-energy scattering state and bound state solutions near s-wave
resonances. For example, it predicts correctly the initial phase
shift as $\delta_1(k=0)=\pi/2$ for scattering state, and a bound
state solution for arbitrarily weak attraction $a_s\rightarrow 0^-$.
However, near p-wave resonances the p-wave pseudo-potential will
produce qualitatively different results compared with exact
solutions. For example, in the limit of $\lambda r_0\ll 1$, the
exact solutions approach free space results, i.e., $\delta_2(k=0)=0$
or $\pi$, and each bound state emerges when $v_p$ goes across a
resonance at certain critical potential depth; on the contrary, the
p-wave model predicts $\delta_2(k=0)=\pi/2$ and a bound state for
any weak p-wave interaction $v_p\rightarrow 0^-$.

Here we analyze the reason why the s-wave pseudo-potential is
approximately valid for SO coupled system while p-wave is not. This
can be explained from the correspondence between the assumptions of
pseudo-potential models and the resulted short-range behavior of
wavefunctions. For s-wave pseudo-potential($\bar{f_0}\neq0,\
\bar{f_1}=0$), the resulted wavefunction does not show $1/r^2$
singularity in p-wave channel, which is consistent with the
assumption of zero scattering amplitude in Eq.\ref{U_P}. However,
the p-wave pseudo-potential($\bar{f_0}=0,\ \bar{f_1}\neq0$) will
induce an additional singularity in s-wave channel, i.e.,
$\psi_0(r)\rightarrow
2\lambda[\frac{\lambda^2+3k^2}{2k}+\frac{\tan\delta}{r}]$ as
$r\rightarrow0$. This in turn requires that the interaction operator
$U$ also generate scattering amplitude in the s-wave channel,
contradictory with the initial assumption that $\bar{f_0}=0$ in
Eq.\ref{U_P}. This paradox also implies that any weak $\bar{f_0}$
will have dramatic interference with the p-wave sector and lead to
qualitatively different results from those using pseudo-potential
entirely in p-wave channel.

The results presented in this section tell us that the original
pseudo-potentials formulated in the absence of SO coupling do not
necessarily apply to the case with SO coupling. The pseudopotential
model will certainly breakdown if the results obtained are
inconsistent with initial assumptions of this model. In this case,
an appropriate interaction model should be constructed in order to
rightly incorporate mixed scatterings between different partial-wave
channels, which is out of the scope of this paper.

\section{Summary}

In this paper, we exactly solve the two-body problem of spin-$1/2$
fermions with isotropic SO coupling under a general short-range
interaction potential, and investigate the validity of s-wave and
p-wave pseudo-potentials formulated in the absence of SO couplings.
Our main results are summarized as follows:

(1)In the presence of isotropic SO coupling, the two-body scattered
wavefunction exhibits exotic dependences on the momentum and phase
shift (Eq.\ref{psi_01_super2}). In each partial-wave channel the
wavefunction at short inter-particle distance is parameterized by
scattering amplitudes of all coupled scattering channels
(Eqs.\ref{short_psi0_f},\ref{short_psi1_f}). This feature reveals
the mixed scattering between different partial-waves that is induced
by the SO coupling.

(2)Under the conditions that the length scale of SO coupling much
longer than the range of the potential($\lambda r_0\ll 1$) and near
s-wave resonances, the s-wave pseudo-potential gives a good
approximation to the low-energy solutions, with the correction
depending on the strength of SO coupling, the finite range of the
potential and contributions from other coupled partial-waves.

(3)Near the p-wave resonance, the p-wave pseudo-potential model
gives low-energy solutions that are qualitatively different from the
exact ones from the square-well potential. The p-wave model alone
can not be applied to the fermion system when the SO coupling
strength is larger or comparable to the Fermi momentum. Its
breakdown is attributed to the inconsistent treatment between the
assumption of the p-wave pseudo-potential and the resulted
short-range singularities of wavefunction in s-wave channel.


Although above results are obtained for the special type of
isotropic SO coupling, they reveal the generic scattering properties
modified by the coupling between spin and orbit. We therefore expect
these results have strong implications to other systems with a
general type of SO coupling.

{\it Acknowledgement.} The author is grateful to T.-L. Ho and H.
Zhai for valuable discussions, and to P. Zhang and S. Zhang for
helpful suggestions. This work is supported by Tsinghua University
Basic Research Young Scholars Program and Initiative Scientific
Research Program and NSFC under Grant No. 11104158.

\appendix

\section{Free-space scattering lengths under a square-well potential}

Far away from the range($r_0$) of the potential, the scattered
wavefunction in $l-$th partial-wave channel reads
\begin{equation}
\psi_l(r)=j_l(kr)-\tan\delta_l n_l(kr)
\end{equation}
here $\delta_l$ is the phase shift which give the scattering length
(with energy-dependence) as
\begin{equation}
a_l(k)=-\frac{1}{r_0^{2l}}\frac{\tan\delta_l}{k^{2l+1}}.\label{al}
\end{equation}
In the limit of $kr_0\ll1$, the effective-range expansion gives
\begin{equation}
\frac{1}{a_l(k)}=\frac{1}{a_l}-\frac{1}{2}r_l k^2,\label{eff-range}
\end{equation}
with $a_l$ the zero-energy scattering length and $r_l$ the effective
range.

For a square-well potential with depth $V_0(<0)$ and range $r_0$, in
the following we scale all lengths ($a_l,r_l$) in the unit of $r_0$
and all momenta ($k,p$) of $1/r_0$; for instance, $\tilde{a}=a/r_0$,
$\tilde{k}=k a_0$. We define two functions as
$\overline{j}_l(x)=x^{-l}j_l(x)$, $\overline{n}_l(x)=x^{l+1}n_l(x)$,
modified respectively from the spherical Bessel and Neumann
functions. We then have
\begin{equation}
\frac{\tan\delta_l}{\tilde{k}^{2l+1}}=\frac{\overline{j}_{l-1}(\tilde{p})\overline{j}_{l}(\tilde{k})-\overline{j}_{l-1}(\tilde{k})\overline{j}_{l}(\tilde{p})}{\overline{j}_{l-1}(\tilde{p})\overline{n}_{l}(\tilde{k})-\overline{n}_{l-1}(\tilde{k})\overline{j}_{l}(\tilde{p})}
\label{delta_SW}
\end{equation}
with $k=\sqrt{mE}$, $p=\sqrt{m(E-V_0)}$. Note that for $l=0$, we
have $j_{-1}=-n_0,\ n_{-1}=j_0$. The zero-energy scattering length
and effective range are given by ($q=\sqrt{-mV_0}$)
\begin{eqnarray}
\tilde{a}_l&=&-\frac{1}{(2l-1)!!(2l+1)!!}\frac{j_{l+1}(\tilde{q})}{j_{l-1}(\tilde{q})},\label{a_l}\\
\tilde{r}_l&=&(2l-1)!!(2l+1)!!\{-\frac{1}{2l-1}+\frac{2l+1}{\tilde{q}^2}\frac{j_{l-1}(\tilde{q})}{j_{l+1}(\tilde{q})}\nonumber\\
&&-\frac{1}{2l+3}(\frac{j_{l-1}(\tilde{q})}{j_{l+1}(\tilde{q})})^2
\}.\label{r_l}
\end{eqnarray}

\section{Scattering under a square-well potential with isotropic SO coupling}\label{SO_SW}

Using the continuity properties of $\psi_0$, $\psi_1$ and their
first-order derivatives at the potential boundary $r=r_0$, we obtain
four coupled equations which can be expressed in a matrix form. For
convenience, we scale all momenta in unit of $1/r_0$ as $\tilde{k}=k
a_0$.

For scattering state ($E=k^2/m>0$), the matrix equation is $A
(1,t,\alpha,\beta)^T=0$ with matrix
\begin{widetext}
\begin{equation}
A=\left(
  \begin{array}{cccc}
    j_0(\widetilde{q}_2) & j_0(\widetilde{q}_1) & j_0(\widetilde{k}_2)-\tan\delta n_0(\widetilde{k}_2) & j_0(\widetilde{k}_1)+\tan\delta n_0(\widetilde{k}_1) \\
    \widetilde{q}_2 j_1(\widetilde{q}_2) & \widetilde{q}_1 j_1(\widetilde{q}_1) & \widetilde{k}_2(j_1(\widetilde{k}_2)-\tan\delta n_1(\widetilde{k}_2)) & \widetilde{k}_1(j_1(\widetilde{k}_1)+\tan\delta n_1(\widetilde{k}_1))\\
    j_1(\widetilde{q}_2) & j_1(\widetilde{q}_1) & j_1(\widetilde{k}_2)-\tan\delta n_1(\widetilde{k}_2) & j_1(\widetilde{k}_1)+\tan\delta n_1(\widetilde{k}_1)\\
    \widetilde{q}_2 j_0(\widetilde{q}_2) & \widetilde{q}_1 j_0(\widetilde{q}_1) & \widetilde{k}_2(j_0(\widetilde{k}_2)-\tan\delta n_0(\widetilde{k}_2)) & \widetilde{k}_1(j_0(\widetilde{k}_1)+\tan\delta n_0(\widetilde{k}_1))\\
  \end{array}
\right)\label{matrix_SW}
\end{equation}
\end{widetext}
here we have used $j_0'(x)=-j_1(x)$,
$j_1'(x)=-\frac{2}{x}j_1(x)+j_0(x)$ to simplify the equations.
$k=\sqrt{mE},\ q=\sqrt{-mV_0}$; $k_2=\lambda+k$, $k_1=\lambda-k$;
$q_2=\lambda+\sqrt{k^2+q^2}$, $q_1=\lambda-\sqrt{k^2+q^2}$.

The zero determinant $|A|=0$ gives rise to two solutions of phase
shift $\delta$. When $\lambda=0$, these two solutions are
respectively resulted from two decoupled equations, and reproduce
the well-known s-wave and p-wave phase shifts in free space as given
by Eq.\ref{delta_SW}.

For bound state, the equations can be obtained straightforwardly by
transformations (see Eq.\ref{bs}) from the equations of scattering
state. The binding energy $E=-\kappa^2/m<0$ can be determined from
the resulted matrix equation $|A_b|=0$. When $\lambda=0$, the
binding energies respectively reproduce the s-wave and p-wave
results without SO coupling.

\end{document}